# Digital Evolution: Novo Nordisk's Shift to Ontology-Based Data Management

Shawn Zheng Kai Tan, Shounak Baksi, Thomas Gade Bjerregaard, Preethi Elangovan, Thrishna Kuttikattu Gopalakrishnan, Darko Hric, Joffrey Joumaa, Beidi Li, Kashif Rabbani, Santhosh Kannan Venkatesan, Joshua Daniel Valdez*, Saritha Vettikunnel Kuriakose*

Novo Nordisk A/S, Novo Nordisk Park 1, 2760 Måløv, Denmark

* Correspondence: jdnv@novonordisk.com, svzk@novonordisk.com

## Abstract

Biomedical data is growing exponentially, and managing it is increasingly challenging. While Findable, Accessible, Interoperable and Reusable (FAIR) data principles provide guidance, their adoption has proven difficult, especially in larger enterprises like pharmaceutical companies. In this manuscript, we describe how we leverage an Ontology-Based Data Management (OBDM) strategy for digital transformation in Novo Nordisk Research & Early Development. Here, we include both our technical blueprint and our approach for organizational change management. We further discuss how such an OBDM ecosystem plays a pivotal role in the organization's digital aspirations for data federation and discovery fuelled by artificial intelligence. Our aim for this paper is to share the lessons learned in order to foster dialogue with parties navigating similar waters while collectively advancing the efforts in the fields of data management, semantics and data driven drug discovery.

## Introduction

The exponential increase in volume, variety, and velocity of biomedical data [1] poses challenges, rendering traditional forms of knowledge management and transfer used in science, like lab notebooks and publications, insufficient. Organisations that can successfully manage their data assets have a significant opportunity in accelerating their drug discovery pipeline[2,3].

Findable, Accessible, Interoperable and Reusable (FAIR) data principles [4] were introduced in 2016, and have since become pervasive in discussions, policies and implementations across disciplines in scientific research.

Organizations have adopted different strategies to ensure FAIR data which can be broadly divided into "FAIR@source" or "born FAIR" [3] and "made FAIR". Most cookbooks and strategies for making data FAIR follow the "made FAIR" strategy where data is "FAIRified" by a dedicated team after it has been generated [5]. In contrast, a FAIR@source strategy empowers data producers to ensure FAIRness of data at point of origin thereby ensuring that the context is accurate and factual rather than inferred. Irrespective of the strategy employed, FAIR data is pivotal in reducing time-to-value and accelerating research [6].

Despite the importance of FAIR principles being recognized widely by the research community, its adoption has proven to be a rather cumbersome task [6]. With the "made FAIR" strategy, lack of access to siloed data and the effort necessary to clean and wrangle data has remained an organizational challenge [7,8]. Whereas, challenges for the FAIR@source approach stem from the scalability requirements for enterprise-level solutions, the substantial investment of time and financial resources into infrastructure, and the technical complexities associated with the implementation process.

A corner stone to ensuring FAIRness of data is rich metadata which ensures context, and as a consequence, improves Findability, Interoperability and Reusability. While controlled vocabularies and taxonomies are widely used in data FAIRification, they are not without their limitations. Ontologies, and data systems that are based on them, overcome many of these limitations, and provide a potential solution to much of the difficulties stated above - they are by design FAIR and scalable. However, ontologies are also challenging to develop, maintain, and implement. This is especially so in a Research & Early Development (R&ED) environment, where success hinges on the use of a variety of tools that are ever changing, which inevitably generates a large variety of data, much of which do not have standard structures. This difficulty is compounded in a company like Novo Nordisk, where 100 years of legacy comes with historical data management systems and practices.

In this manuscript, we describe our strategy for digital transformation in Novo Nordisk R&ED using an Ontology-Based Data Management (OBDM) strategy for bridging the gap between data producers and consumers to ensure the best use and reuse of our data assets. We consider this publication to be among the initial discussions addressing a FAIR@source strategy and our hope for this manuscript is to offer valuable insights for FAIR practitioners and developers, shedding light on the challenges encountered by large-scale pharmaceutical companies, and by extension, various data-producing corporations. Furthermore, we aim to share the lessons learned through these experiences, fostering a broader dialogue with parties navigating similar waters while collectively advancing the efforts in the fields of semantics and data management.

## Results

### An Ontology-Based Data Management strategy

In Novo Nordisk R&ED, we are implementing a comprehensive "FAIR@source and scale" approach to data management. The FAIR@source strategy significantly expedites the response time to scientific queries while ensuring factual context and reducing time-to-value. Simultaneously, our emphasis on FAIR@scale prioritizes scalability and the development of enterprise-level solutions. This strategic emphasis is crucial for the practical utilization of solutions within a large organization, a prerequisite particularly pertinent to global organizations that span multiple geographies. Ensuring the scalability of our data management solutions is integral to their effective implementation and utility across diverse facets of our organization.

To operationalise our "FAIR@source and scale" ambitions, we employ an OBDM strategy. In this strategy, centralized ontologies serve as the Single Source of Truth (SSOT), ensuring consistency and accuracy in data representation. Employing ontology-based structures aids in the alignment and integration of data across various domains, thus streamlining the process of adhering to the FAIR principles from the point of data creation. This approach also ensures scalability within the operational framework of Novo Nordisk.

### Building an Ontology-Based Data Management ecosystem

Having a good approach for ontology consumption is crucial for the development of an OBDM ecosystem. As of January 2024, BioPortal [9] contains 1065 published models, the Ontology Lookup Service [10] by EMBL-EBI and Ontobee [11] contain 246 and 263 ontologies respectively. Given the large corpus of work that already exist, one should, as far as possible, utilise existing ontologies instead of creating a new reference ontologies. In the spirit of FAIR principles, reusing ontologies would lead to greater interoperability. The choices for ontologies have been extensively described in literature [12,13]. Regardless of the choice of public ontology, it is likely that none of them are fully fit for purpose for the organisation. This is not surprising as most public ontologies are built as general-purpose reference ontologies, while organisational requirements tend to be specific. In order to cater to our specific requirements, our OBDM strategy advocates for the development of organisation-specific ontologies derived from public ontologies. This approach allows for the flexibility required to cater to the needs of the organisation, while remaining interoperable with external data. We ensure that concepts specific to our organisation are parented by a public ontology which allows for easier interoperability, a strategy also used by ontology extensions [14].

While various strategies exist for organizations to bridge the gap between the scope of public ontologies and organisational needs, our approach was shaped by several key considerations. Through multiple iterations, we have been able to continuously learn and refine our approach. Despite the multitude of ontologies and associated resources, it became apparent that no singular representation could fully satisfy our needs across all domains of interest. With this in mind, we developed what we term "domain models" in which we defined the domains which are of interest, identified relevant ontologies, and created our own internal taxonomies based on them (described below). Another key decision was based on the OBO Foundry principle [12,15] of orthogonality which asserts that for each domain there should be convergence upon a single ontology. Based on this, we decided against bringing in multiple full ontologies with the same scope since having multiple concepts with the same definitions would lead to difficulties in integration.

Our domain models are based either directly on a public ontology or on a composite of multiple ontologies. The latter are amalgamated to form a cohesive model through a process that involves extracting subgraphs and merging them where appropriate. This process also includes the harmonization and merging of concepts from multiple ontologies within the same domain. To ensure flexibility, we mint new URIs for our domain models. This allows us to modify logical axiomatization of ontologies or append ontology terms to other ontologies. We use Simple Standard for Sharing Ontological Mappings (SSSOM) [16] systems to maintain interoperability with external sources, allowing us to update our internal ontologies, and keep in sync with public ontologies, avoiding drift. This was important to us as ontologies are not static artefacts, but models that evolve alongside knowledge. Additional benefits of using shared standards include easier utilization of community efforts like biomappings [17], and availability of open source tooling (eg sssom-py). From here, tools like the aforementioned biomappings and OXO [18] can help naturalise annotated external data to our ontologies. Additionally, semiautomated systems using named entity recognition (NER) can aid in the annotation of unannotated data.

As our upper ontology, we use BFO [19] which allows us to use reasoner-based coherency checks in the future. Deciding on an implementation for a middle/unifying ontology is a bit more complicated. For example, in the biomedical area, the OBO foundry has created an experimental unifying middle ontology, Core Ontology for Biology and Biomedicine (COB)(https://github.com/OBOFoundry/COB), that aims to bring together key terms from OBO ontologies. Work is also underway by Pistoia Alliance to develop a similar middle/upper ontology to unify high level concepts in the pharma space (termed Pharma General Ontology (PGO)) and we are actively contributing to the thought leadership underlying its construction. Federated solutions such as mapping of terms is an alternative to having unifying ontologies, and community efforts like biomappings [17] are already ongoing. However, since these unified solutions are in their infancy, we decided to take an approach interoperable with either of them. As of April 2024, we do not map to any middle ontology, but instead directly to BFO, with the knowledge that mappings can be made in the future.

## Securing interoperable scientific metadata

Our FAIR@source and scale strategy relies on metadata in all applications being interoperable and served from an SSOT. We use taxonomies derived from ontologies to act as the SSOT for scientific metadata. Since ontologies are complex and difficult to maintain, we build SKOS taxonomies based on public ontologies to maintain enrichments (terms specific to Novo Nordisk) to domain models. These taxonomies only maintain annotations, hierarchical structures, and minimal relationships between concepts (as opposed to ontologies, which contain a richer and more expressive relationships). This allows us a quick turnaround required in an R&ED environment. Given that our enrichments are always parented by a concept that is derived from a public ontology, it affords us flexibility in our system. For example, if we decide to maintain owl ontologies at a later date, our enrichments are already parented by concepts modelled as such, and conversion of enrichment to owl objects can be as rich or shallow as we choose. We however do acknowledge that this comes with a huge risk of drift that may lead to issues down the road [20]. To mitigate this, we ensure that conversions are done in automated pipelines which allow us to easily update to new versions of ontologies – something we do on a regular basis. Delivery of these taxonomies to stakeholders takes the form of ontology governed controlled vocabularies (a structured list of concepts derived from the taxonomies) delivered in any form the stakeholders prefer – mostly APIs.

## Integration across data silos using knowledge graphs

Given the legacy of a century-old organisation, we have diverse data sources originating from legacy and federated systems. Ensuring semantic interoperability across these silos gives us the opportunity to better leverage insights across the data landscape. One way to enable semantic interoperability is through the use of a Knowledge Graph (KG). However, the ability to integrate disparate data sources into a KG can present a formidable challenge. An effective solution to this issue can be found in the implementation of a Virtual Knowledge Graph (VKG) or Ontology-Based Data Access (OBDA) approach. In this approach, data sources such as databases are mapped to an ontology, thereby presenting a unified KG [21,22]. This methodology offers significant advantages, including the ability to leverage existing security measures and access controls, as the data remains in its original location and does not require materialization. Furthermore, a VKG approach provides a more scalable method for data ingestion. Maintaining scalable mappings, as opposed to the resource-intensive processes of data materialization and constant reindexing, results in a more efficient and sustainable system. In essence, virtualization in the context of a semantic KG offers a flexible and efficient approach to data integration, allowing for the addition or removal of data sources without disrupting the overall structure of the KG.

Developing a KG takes a lot of time, resources, and commitment. Legacy and disparate sources of data, for example, require huge curation effort. We therefore have adopted the strategy of incremental improvements based on concrete uses cases with stakeholders that can champion it. Our KG is built with scalability, useability, and flexibility in mind. Given that we link our data through our own internal URI, any rewiring needed can be done easily. Changing ontologies can be done simply by mapping our internal URIs to the new ontology concepts' URIs. Furthermore, if we decide to eventually maintain our own internal owl ontologies, switching can be done in very similar ways. Public ontologies are also brought in as imports in a modular fashion, this would mean that if we require slices of the ontology/KG for specific future applications, it can easily be done. All this points to a generalisable, flexible, and scalable system that fits our strategy of incremental improvements.

We take a semantic approach to our KG construction with public ontologies as its underlying structure. Our semantic KG utilises our domain models' links to public ontologies to establish a bridge between internal URIs and those of public ontologies using equivalence assertions. Figure 1 shows a diagrammatic representation of how we built our application-specific ontology. In this example, the aim was to query the graph on any "experiment" that "involves" a "chemical entity" that has_role "anti-obesity agent". The solid lines show explicit relationships between entities in our graph, while the dotted lines show inferred relationships. The inferred response to the query above is highlighted in red.

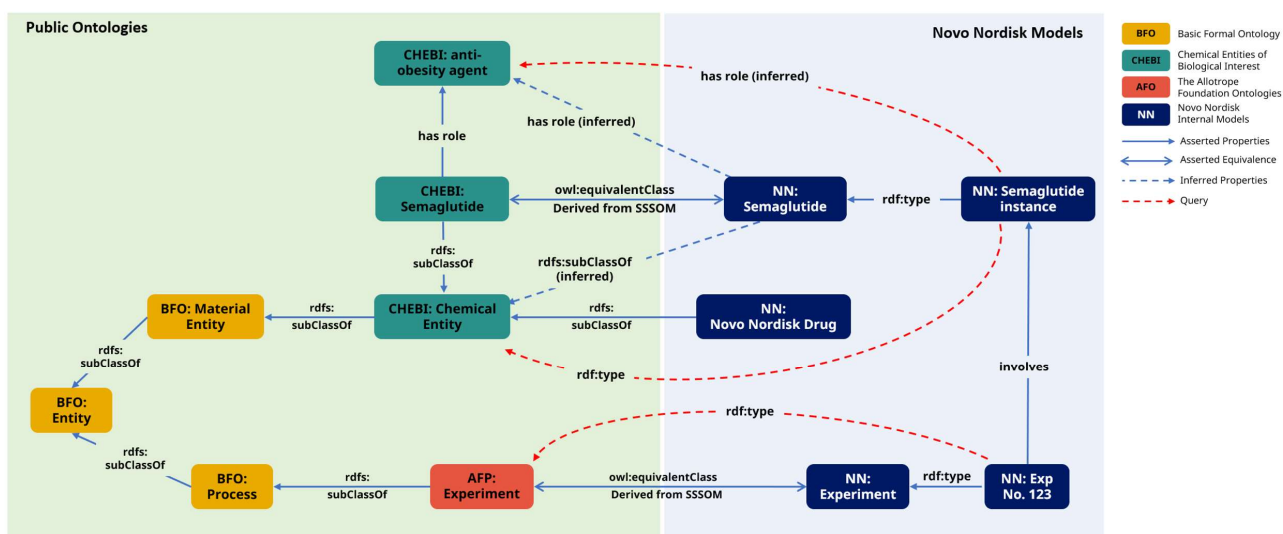

**Figure 1.** Diagrammatic representation of our knowledge graph approach which illustrates the response path to the query: any "experiment" that "involves" a "chemical entity" that has_role "anti-obesity agent". The left side represents public ontologies with BFO (yellow) as the upper ontology, and CHEBI (green) and AFO (red) as ontologies of interest. Novo Nordisk specific nodes (blue) are linked either by

owl:equivalentClass or rdfs:subClassOf. The solid lines show explicit relationships between entities in our graph, while the dotted lines show inferred relationships. The inferred response to the query is highlighted in red.

## Discussion

In this manuscript, we have described our strategy for building a FAIR@source and scale ecosystem based on OBDM. Although such a strategy has its merits, initiating it is notoriously challenging. Our four-year journey has underscored the importance of addressing the immediate needs of stakeholders while also planning for future use cases. Key components of an OBDM strategy include the models themselves and the capacity of downstream applications to utilize these models from a centralized source. Thus, the complexity of implementing this strategy is multifaceted and heavily reliant on collaboration across various sectors of a large organization. Inevitably, this raises the question of the value gained from investing in the development of such intricate systems, especially given the complexity, time, and costs involved. Developing an OBDM ecosystem is a strategic investment that can yield significant benefits, particularly in the realms of decentralized data architecture, semantic interoperability, and consequently, Artificial Intelligence (AI). Specifically, the integration of ontologies into modern AI architectures can considerably enhance data interpretation and utilization.

Ontologies provide a semantic context that enhances the capabilities of AI technologies, such as Large Language Models (LLMs), machine learning algorithms, and knowledge graph-based systems. Consider the application of an LLM within the context of a drug discovery workflow. Without a systematized semantic context, the LLM may encounter difficulties in interpreting the intricate relationships between various biological entities or in disambiguating between similar entities (e.g. diseases, rare diseases, and symptoms or signs [23]). The integration of biomedical ontologies, which also function as curated knowledge bases, such as the Gene Ontology (GO) [24,25] or the Chemical Entities of Biological Interest (ChEBI)[26], can substantially mitigate these challenges. These ontologies provide a semantic framework that equips the LLM with the necessary tools to accurately interpret complex biomedical data, thereby enhancing its accuracy and reliability.

One of the biggest challenges with the use of LLMs has been hallucinations, a common and dangerous occurrence related to the way these models operate. This highlights the need for rigorous validation processes to address these issues [25-27]. Ontologies, and semantic knowledge graphs developed from them, can function as a 'reality check' for LLMs, ensuring that outputs are both semantically and contextually accurate. An example of how this can be done is through the use of the Common Coordinate Framework

(CCF) validation tool (https://github.com/hubmapconsortium/ccf-validation-tools), which utilizes ubergraph to validate structured expert-curated tables and atlases. Such tools can be utilized to validate the accuracy of the information generated by LLMs. Additionally, the incorporation of KGs into LLMs enhances their performance by enabling Retrieval Augmented Generation (RAG) architectures, affording the ability to leverage search for information from federated sources while responding to user queries [30]. This integration, exemplified by efforts such as the Monarch Initiative [31], has demonstrated improved language model performance within the biomedical domain [32]. Generative AI can further be used to democratize the use of KGs by enabling the generation of querying languages such as SPARQL using natural language, enhancing the accessibility and usability of KGs [33]. Overall, ontologies and semantic knowledge graphs play a pivotal role in reducing hallucinations and enhancing the performance of mixed generative retrieval strategies.

Beyond the advantages that an OBDM strategy brings to the AI capabilities of an organization, perhaps most notably, this approach plays a crucial role in the implementation of enterprise search. The pharmaceutical industry generates a vast array of complex and heterogeneous data, spanning from chemical and biological data to clinical trial data and patient records. Often, this data is stored in disparate systems and in various formats, posing a challenge when it comes to efficiently searching and retrieving relevant information. An OBDM strategy addresses this challenge by providing a unified view of the data, facilitating the integration of data from different sources and in different formats, thereby breaking down the silos and enhancing the accessibility and interoperability of the data. In the context of enterprise search, this means that users can search for information across different systems using a common set of terms and receive results that are contextually relevant and comprehensive. For instance, a researcher looking for information on a specific drug compound can use the same search terms to retrieve information from chemical databases, clinical trial databases, and patient records within the confines of our organization's access policies. Furthermore, the inherent structural semantics of ontologies can improve the specificity and relevance of search results. This is accomplished by leveraging the axiomatic and hierarchical relationship context between various data elements, moving beyond the limits of string-matching techniques. By providing a unified and semantically rich view of the data, an OBDM approach supports more informed decision-making.

Finally, due to the diverse nature of data systems, the presence of a decentralized data architecture becomes essential for ensuring appropriate data ownership and governance. This approach fosters a scalable and flexible method for managing data, enabling individual teams to effectively utilize and oversee their specific data assets. A Data Mesh paradigm [27] fulfils such a need through the distribution of data across distinct domains, each characterized by its unique data product and product owner. This distribution, while advantageous in certain aspects, can introduce complexities in data integration and interoperability due to the inherent heterogeneity of data across the domains. An OBDM strategy effectively navigates these complexities by providing a unified semantic framework enabling seamless data integration across the different domains within the Data Mesh. We see the OBDM enabling the federation of data in Novo Nordisk by serving as a semantic mediator that harmonizes disparate data sources. In this context, ontology plays a pivotal role in providing a shared and common understanding of the data domain. This integration ensures that data from disparate domains can be coherently understood and leveraged for various applications in a unified manner, thereby overcoming many of the challenges posed by a distributed system. Moreover, this approach actualizes the full potential of a decentralized data architecture by enhancing data accessibility, interoperability, and usability. The OBDM strategy employs a global-as-view (GAV) strategy, where the data sources are mapped to ontologies. This ensures that data queries can be executed across multiple databases in a semantically consistent manner, thereby improving the efficiency and accuracy of data retrieval. Furthermore, the semantic relationships encoded in the ontology can be leveraged to infer new knowledge from federated data. This not only enriches the data exploration and discovery process but also enhances the expressivity and reasoning capabilities of the data federation system. The OBDM's ability to handle implicit semantics and ontological inconsistencies further strengthens its role in data federation. By resolving semantic conflicts and ambiguities, OBDM ensures the integrity and reliability of the federated data.

In conclusion, while FAIR principles have been codified, their execution has proven difficult, especially at an enterprise level in a heterogenous fast paced environment like pharma R&ED. To address this, we implemented an OBDM strategy, as discussed in this manuscript, which not only addresses our need for scale but also ensures that the data is FAIR@source. An added advantage from such an ecosystem is that it reduces time to value and accelerates data driven research and decision making. In this communication we focussed on the technical implementation of our preferred approach, but the effort required to make this happen both from a change management perspective and resource required to bring scientific knowledge to ontologists should not be discounted. Stewardship is an essential component of this playbook facilitating coordination between research scientists, data scientists, and semantic experts. Additionally, an organisational commitment at different levels starting from leadership commitment to state of the art infrastructure to bench scientists willing to collaborate to keep our knowledge bases up to date is a must. Along our four-year transformation journey, we were helped in our ambitions by the progress in technology, specifically LLMs with the consequent increased attention to semantics. The rapid progress served to underscore the need for a sustainable, scalable, and flexible data foundation which we believe is addressed by our OBDM strategy. We hope that by sharing our journey and technical blueprint, we can foster dialogue, exchange learnings and address potential pain points. Despite the challenges we encountered along the way, we believe that the value that an OBDM ecosystem brings outweighs the challenges, and therefore is worth investing in.

## Methods

### Consuming public ontologies

As of April 2024, we consume 13 ontologies, either in whole or in part, to develop our 13 domain models : uberon [28], cell ontology (CL) [29], Cellosaurus (CVCL) [30], bioassay ontology (BAO) [31], ontology for biomedical investigations (OBI) [32], allotrope foundation ontology (AFO) [33], gene ontology (GO) [24,25], protein ontology (PR) [34], Chemical Entities of Biological Interest (CHEBI) [26], Mondo Disease Ontology (MONDO) [35], human phenotype ontology (HPO) [36], NCBITaxon [37], and Quantities, Units, Dimensions and Data Types Ontologies (QUDT) [38]. These ontologies were selected based on a combination of need/use cases, and guidelines from Pistoia Alliance [39] and OBO Foundry [12]. Figure 2 shows a diagram of the public ontologies we use and the domain models they fuel.

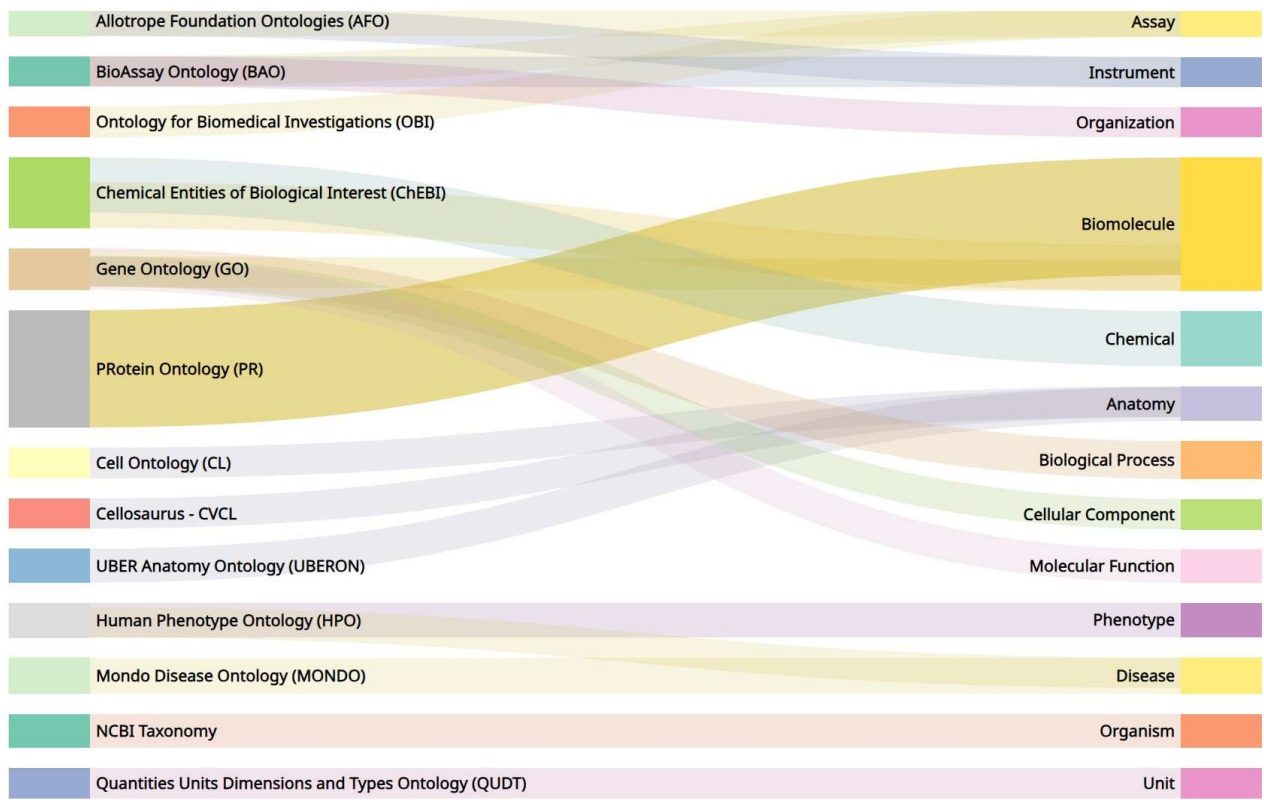

**Figure 2.** Sankey diagram of the public ontologies we use (on the left) and the domain models (on the right) they are used by. While some public ontologies directly contribute to our domain models, most of our domain models are composites of different public models or parts thereof.

### Conversion to SKOS taxonomies

In order to convert OWL ontologies to SKOS taxonomies, we developed custom scripts for each ontology. This allows us the flexibility to ensure coherency between taxonomies converted from different ontologies. The input files to these convertors vary depending on the ontology. The convertor scripts are a series of SPARQL queries which convert owl:Class to skos:Concept and rdfs:subClassOf assertions to skos:broader and pull over selected annotations (eg labels, textual definition, etc.). Owl EquivalenceAxioms are relaxed and treated similarly to rdfs:subClassOf. During the conversion, new URIs are also minted, and the convertor outputs an SSSOM file with skos:exactMatch as the predicate_id between the new URIs and the public URIs.

### Enrichment to taxonomies

Enrichments to taxonomies are added as skos:narrower concepts of concepts derived from a public ontology. This is done through a centralized taxonomy management system either manually or through a proprietary templating system. Regardless, quality of all enrichments is ensured through internal review process.

### Delivery of controlled vocabularies

Controlled vocabularies are delivered to stakeholders in the form of curated dropdowns that are built in collaboration with users. We make them available via APIs from our taxonomy management system. In order to obtain terms in a given dropdown, our stakeholder's system issues pre-determined API call to our taxonomy management system which returns the terms in the requested dropdown. This eliminates the need for manual importing or updating by stakeholders, saving time and minimizing errors.

### Building a knowledge graph application specific ontology

In order to construct the ontology that underpins our KGs, we leverage the links in our taxonomies to public ontologies. These links are maintained according to SSSOM standards. We use a SPARQL query to link these concepts using equivalence assertions. To build an application ontology that underlies the KG, we perform induced subsetting from public ontologies using ROBOT [40], only bringing in the concepts we need using a Syntactic Locality Module Extractor (SLME) method. We link the subset public ontologies using BFO (denoted in yellow nodes in Fig 1). The above described steps are designed as a scalable workflow which mimics the dynamic import system of the ontology development kit (ODK) [41] but removes the need for Docker.

## Conflict of Interest

All authors are full time employees of Novo Nordisk. TZKS is part of the OBO Foundry Operations Committee. JDNV and SVZK are steering committee members for the PGO project in the Pistoia Alliance.